# Who's Ditching the Bus?

Simon J. Berrebi[a,], Kari E. Watkins[a]

[a]*Georgia Institute of Technology*
*School of Civil and Environmental Engineering*

**Abstract**

This paper uses stop-level passenger count data in four cities to understand the nation-wide bus ridership decline between 2012 and 2018. The local characteristics associated with ridership change are evaluated in Portland, Miami, Minneapolis/St-Paul, and Atlanta. Poisson models explain ridership as a cross-section and the change thereof as a panel. While controlling for the change in frequency, jobs, and population, the correlation with local socio-demographic characteristics are investigated using data from the American Community Survey. The effect of changing neighborhood demographics on bus ridership are modeled using Longitudinal Employer-Household Dynamics data. At a point in time, neighborhoods with high proportions of non-white, carless, and most significantly, high-school-educated residents are the most likely to have high ridership. Over time, white neighborhoods are losing the most ridership across all four cities. In Miami and Atlanta, places with high concentrations of residents with college education and without access to a car also lose ridership at a faster rate. In Minneapolis/St-Paul, the proportion of college-educated residents is linked to ridership gain. The sign and significance of these results remain consistent even when controlling for intra-urban migration. Although bus ridership is declining across neighborhood characteristics, these results suggest that the underlying cause of bus ridership decline must be primarily affecting the travel behavior of white bus riders.

## 1. Introduction

In 2018, bus ridership in the United States fell to its lowest point since 1973 following six years of consecutive decline (Dickens and MacPherson, 2019). Between 2012 and 2018, vehicle miles traveled have increased every year and reached their highest point ever (Federal Highway Administration, 2019). The declining bus ridership, therefore, feeds into low-occupancy driving, thereby contributing to congestion, pollution, and traffic fatalities. Meanwhile, the lost fare revenue makes it more difficult for transit agencies to deliver service. To break this downward spiral, practitioners and policy-makers need to understand the underlying causes of bus ridership decline.

The causes of this decline remain a mystery. Factors that are widely recognized in the literature as positive contributors to bus ridership, including service levels and urban population have

---

*Email address:* `simon@berrebi.net` (Simon J. Berrebi )

increased in recent years (Kain and Liu, 1999; Taylor et al., 2009; Chen et al., 2011; Boisjoly et al., 2018). Following post-recession service cuts, transit agencies in the United States have increased bus service (vehicle revenue miles) by 5.80% between 2012 and 2018 (Federal Transit Administration, 2019). Meanwhile, urban core population has grown every year since 2006 (Frey, 2018). Finally, studies on the impact of ridehailing services on transit ridership have reached different conclusions (Boisjoly et al., 2018; Graehler Jr et al., 2019; Hall et al., 2018). To date, virtually all the research investigating ridership trends over time have used the metropolitan area or transit agency as the spatial unit of analysis. However, the dynamics affecting bus ridership trends are likely to be taking place at a far more disaggregated scale.

In order to understand **why** bus ridership is declining, we must first ask **who**. Transit agencies serve diverse constituencies, which have different travel behaviors. On an individual level, the demand for buses changes over time in reaction to personal circumstances, service availability, and competition from alternative modes. The external factors driving travel demand and the reaction they elicit are not homogeneously spread across socio-demographic groups (Lu and Pas, 1999). For example, the rise in telecommuting almost only applies to white-collar jobs. Another example is dropping car ownership costs, which affect everyone but have more impact on the travel behavior of people with less disposable income. Therefore, whatever is causing the bus ridership decline on a national level may have different effects on individual patrons.

Identifying which passenger characteristics are most closely associated with the decline is a necessary step towards understanding the causes of behavior change. If the ridership loss is particularly prevalent among certain socio-demographic groups, then the factors leading to this shift in travel behavior could be identified. If, however, the ridership decline is evenly distributed, then the underlying cause could be a blanket factor affecting all in the same way. Based on this determination, transit agencies would then be able to anticipate future changes in bus ridership and find ways to reverse the trend. They could, for example, decide whether to lure back lost riders with more service or concentrate on other promising market segments.

The socio-demographic characteristics associated with bus ridership change could not be fully understood without taking into consideration intra-urban migration. When people who depend on transit as their main mode of transportation can no longer afford to live close to frequent bus service, they are likely to relocate in lower-density suburbs with less access to transit (Hwang and Lin, 2016; Brown-Saracino, 2017). To control for the potential impact of gentrification on ridership, we consider the shifting demographics in our model. In this paper, the word "gentrification" is quantified in terms of local demographic shifts regardless of prior demographic composition.

We use fixed-effects models to addressed the endogeneity between travel demand and supply; each individual location serves as its own control for the variation taking place over time (Berrebi et al., 2020). While seeking to identify the neighborhood socio-demographics associated with ridership and change thereof, we control for higher-order effects of population, jobs, and service frequency. The neighborhood characteristics correlated with bus ridership at one point in time and with the change in bus ridership between 2012 and 2018 are modeled with fixed-effects Poisson models. This time period corresponds to the nation-wide bus ridership decline. This analysis is based on four transit agencies, which were selected due to the high-quality of their passenger-count data:

- Tri-County Metropolitan Transportation District of Oregon (TriMet) in Portland OR
- Miami-Dade Transit in Miami, FL
- Metro Transit in Minneapolis/St-Paul, MN
- Metropolitan Atlanta Rapid Transit Authority (MARTA) in Atlanta, GA

This paper proceeds as follows: the next section describes the research done thus far on who's ditching the bus. The Data section introduces the model variables and their sources. The Case Study section introduces the four transit agencies featured in this research. The Model section describes the framework to determine the relationship between explanatory variables and bus ridership. Ridership is modeled as a cross-section and as a panel in the two following sections. The impact of shifting socio-demographics is then evaluated. Finally, the broader implications of this research are discussed in the conclusion.

## 2. Literature Review

Due to a long history of deliberate policies, US cities remain largely segregated by race and class (Squires and Kubrin, 2005; Rothstein, 2017). While the identity of local residents does not directly impact travel behavior, it is connected to important factors such as job location, working hours, parental obligations, and others that are not entirely measurable (Lu and Pas, 1999). Neighborhood socio-demographics are also connected to accessibility, which factors into all aspects of travel behavior, at the trip, mode, and route level (Levinson, 1998; Grengs, 2001; Manaugh et al., 2010). This is why neighborhood socio-demographics have long been used in the literature to understand changing travel patterns. In the next subsection, research considering static characteristics at the local and individual level is described. The following subsection reviews studies modeling the effects of intercity migration.

### 2.1. Changing Demand for Transit

To capture the relationship between local socio-demographics and ridership, studies have crossed stop-level ridership data with the American Community Survey. Dill et al. (2013) model stop-level bus and rail ridership in Portland, Eugene, and Medford, OR. They find that across the three cities, the proportion of local residents who are white, college-educated, and have access to a car correlate negatively with ridership. Frei and Mahmassani (2013) evaluate how static socio-demographics explain stop-level ridership. They find that the proportion of residents who are employed and under 17 years-old correlates with higher ridership. Mucci and Erhardt (2018) model factors affecting stop-level ridership in 2009 and measure the change in these factors between 2009 and 2016. Assuming that the effect of income on ridership has remained constant, they extrapolate that the 6% increase in high-income households could be responsible for 4% of the decrease in ridership. Manville et al. (2018) use the California Household Travel Survey to model the change in ridership at the regional level. They find that vehicle ownership is negatively correlated with ridership.

Several studies have used surveys to establish the connection between ridership and socio-demographics at one point in time. Based on the Civic Census in Calgary, Canada, Pasha et al. (2016) find that income is negatively correlated to transit mode share at the neighborhood level. Based on the 1995 National Household Travel Survey (NHTS), Giuliano (2005) find that low-income earners and African-Americans are more likely to have less access to private vehicles and to use public transportation more often. While these studies help understand who rides the bus at one point in time, they do not explain changes in travel behavior over time.

Changing transit mode-share both within and between age cohorts has been evaluated in several studies. Based on travel surveys in Montreal from 1998, 2003, and 2008, Grimsrud and El-Geneidy (2013) and Grimsrud and El-Geneidy (2014) find that transit mode share declines over the course of one's life due to life cycle factors such as commute distance, parental responsibilities, etc. However, when controlling for these variables, people in their 20s exhibit greater transit mode shares than previous cohorts. Brown et al. (2016) use a similar methodology, comparing the 2001 and 2009 NHTS. They find that the greater propensity for young adults to ride transit than older cohorts can be attributed to life cycle factors, which are common to young people but unlikely to persist as they age. Coogan et al. (2018) compare the 1990 and 2009 NHTS, but do not control for life cycle factors. They find that people over 25 years-old are making more transit trips than previous cohorts at the same age. The ridership of 16 to 24 years-old is not significantly different than previous cohorts. While the shifting demand for transit within age-cohorts over time has been addressed in the literature, there is a gap in our understanding of the changing travel behavior in other socio-demographic groups.

2.2. Urban Migration

A body of research has quantified the impact of socio-demographic change on ridership change over time at the agency or regional level. Driscoll et al. (2018) estimate the effects of population aging on ridership. Assuming that travel demand and mode share by age bracket have remained constant since 1980, they estimate that 3% of the 20% bus ridership decline could be due to aging population. In a longitudinal study of Utah Transit Authority ridership over 10 years, Lyons et al. (2017) find that the change in proportion of white residents at a regional-level correlates negatively with the change in ridership. Boisjoly et al. (2018) model the factors affecting ridership change in 25 North American transit agencies over 14 years. They find that the change in proportion of carless households is positively correlated with ridership change. In a series of scatter-plots, Watkins et al. (2019) also find a positive relationship between the change in proportion of carless households and the change in ridership between 2012 and 2016 for clusters of large metropolitan areas but not for smaller ones. These results help understand how the evolution of transit constituencies and their needs affect ridership on an aggregate level, but they lack the granularity to capture the impact of migration patterns happening locally. In particular, as people who depend on transit move further from city centers, ridership may be affected.

The term "gentrification" was first coined over 50 years ago by Glass (1964) to describe the influx of a "gentry" in lower-income neighborhoods, although the precise definition remains a point of debate among social scientists and economists (Brown-Saracino, 2017). There is a consensus in the literature that gentrification alters the socio-demographic composition of neighborhoods through attrition rather than direct displacement (Slater, 2009; McKinnish et al., 2010; Ellen and O'Regan, 2011; Brown-Saracino, 2017). Low-income and minority residents are likely to relocate often regardless of gentrification (Ihrke et al., 2011; Phinney, 2013). As rents increase, these residents are unable to move back into the neighborhood, leading to gradual socio-demographic change (Guerrieri et al., 2013; Freeman, 2005; Hyra, 2016).

The rise of gentrification coincides with the recent nation-wide ridership decline in time and space; there has been an upsurge since 2000 (Freeman and Cai, 2015; Maciag, 2015; Zook et al., 2017), especially in places where transit is most available, i.e. dense, poor neighborhoods located near the central business district in mid-sized to large cities (McKinnish et al., 2010; Hwang and Lin, 2016; Brown-Saracino, 2017; Richardson et al., 2019). Meanwhile, suburbs have grown increasingly poor and diverse since 2000 (Kneebone and Holmes, 2014; Kneebone, 2017). While evidence from Glaeser et al. (2008) and Pathak et al. (2017) suggests that low-income residents tend to relocate close to transit service both in urban cores and in the suburbs, the lower service frequencies and vehicle ownership costs in the suburbs could be contributing to a decline in individual demand for transit. In an analysis of metropolitan area-level ridership change between 1990 and 2000, Thompson and Brown (2006) found that the increasing proportion of Black population was linked to ridership increase in small metropolitan areas (between 0.5 to 1 million) but not larger regions. More recently, Higashide and Buchanan (2019) found in surveys and focus groups that low-income residents were more likely to move in search of cheaper housing and experienced the greatest loss of transit access between 2016 and 2018. There still lacks, however, empirical research connecting the change in bus ridership over time with the change in neighborhood socio-demographics.

*2.3. Conceptual Framework*

This paper makes three main contributions. The first is to verify results from the literature regarding socio-demographic characteristics associated with bus ridership at one point in time. The second, which is by far the most important, is to evaluate the shifts in travel behavior within socio-demographic groups with the goal of furthering our understanding of the current ridership crisis. We are particularly interested in exploring how the variables that correlate with high-ridership at a point in time, including race, vehicle ownership, and education, help explain the ridership change over time. The third contribution of this research is to evaluate how much socio-demographics surrounding bus service have shifted over time and whether urban migration could explain some of the ridership decline on a hyper-local level.

## 3. Data

In this paper, the spatial unit of analysis is the route-segment. Route-segments are clusters of seven adjacent stops on the same route and direction. Only stops that were neither added nor removed between 2012 and 2018 were considered when constructing route-segments. The last several stops located at the end of a route were included in the route-segment immediately upstream. Bus ridership represents 58.4% of overall transit ridership in Portland, 62.7% in Miami, 68.1% in Minneapolis/St-Paul, and 44.1% in Atlanta. There were no new rail openings between 2012 and 2018 in Miami and Atlanta. In Minneapolis, where the Green light-rail line opened in 2014 and the Bus Rapid Transit A line opened in 2016, the two bus routes running parallel, 16 and 84, were excluded from the analysis. In Portland, no bus route runs parallel to the Orange light-rail line, which opened in 2015.

The total ridership at each segment was divided by the number of stops. A 1/4 mile buffer surrounding each segment, which corresponds to the typical walking distance to bus stops according to (Kittelson Associate, 2013, §5-10), was drawn and overlaid with population, jobs, and socio-demographic data. Blocks and Block Groups that fell partially inside route-segment overlays were allocated proportionally to the overlapping area assuming uniform density. All the terms and variables are defined in Table 1. People and jobs within walking distance of several stops in the same segment were only counted once. They could, however, be assigned to multiple segments.

Ridership data were obtained from the four transit agencies featured in this study. The data came from Automated Passenger Counters (APC), which are beacons aboard the buses recording passenger activity at each stop. Each transit agency provided the research team with average passenger boardings and alightings per stop per trip per markup[1]. The average boardings and alightings were added across trips to obtain average weekday ridership.

Frequency data were obtained from the General Transit Feed Specification (GTFS), a transit schedule metadata format for stops, routes, and schedules. The number of daily weekday trips served as an explanatory variable. We also used the GTFS to cross-check the completeness of APC data. Stops that had more than one scheduled trip missing from APC were removed from the analysis. This step helped ensure that changes in bus ridership could not be attributed to missing data.

Static socio-demographic data, which represent a single time-period, were obtained from the American Community Survey (ACS). The American Community Survey is the largest household survey administered by the Census Bureau with an annual sample size of 3.5 million addresses (Graham et al., 2014). Neighborhood socio-demographics are available at the Block Group level on a five-year rolling basis. In this study, we used the latest ACS data available, which were collected between 2013 and 2017. Measuring the change in population, jobs, and socio-demographics would require at least two non-overlapping five-year estimates. Since we do not have ten years of reliable APC data for the agencies in this study, ACS data can only be used to measure static characteristics.

---

[1] A markup is a 3-4 month period of schedule validity.

The *change* in local population, jobs, and socio-demographics were obtained from the Longitudinal Employer-Household Dynamics (LEHD). The LEHD are data products compiled by states using unemployment insurance earnings and published by the US Census Bureau. The number of residents and jobs are provided at the Census Block level, by year. The data are available on the 2010 Census boundaries between 2011 and 2015. We projected the change in population, jobs, and socio-demographic data to continue on the same trajectory. In other words, the yearly rate of change was assumed to be the same between 2012 and 2018 as between 2011 and 2015.

The LEHD data have several drawbacks compared to ACS. Unlike ACS, which strives to survey a fully representative sample of the population, only on-the-books workers are represented in LEHD. Students, retirees, unemployed people are not counted. Since the data are based on unemployment insurance, undocumented workers are also unrepresented (Andersson et al., 2014). On the other hand, individuals working several jobs can be counted multiple times. Due to these limitations in the LEHD data source, we only use the data to measure the change in population, jobs, and socio-demographics over time. We rely on ACS to account for static characteristics.

Table 1: Summary of variable definitions

| Variable | Definition |
| --- | --- |
| Rid | Total weekday passenger boardings and alightings |
| Freq | Total weekday vehicle-trips |
| Pop+Job | Total population and jobs within $\frac{1}{4}$ mi of segment |
| $\text{Dem}_{ZeroVehHH}$ | Proportion of households with zero vehicles |
| $\text{Dem}_{White}$ | Proportion of residents who are white (both non-Hispanic and Hispanic) [2] |
| $\text{Dem}_{HighSch}$ | Proportion of residents who completed high-school or less |
| $\text{Dem}_{Millennial}$ | Proportion of residents aged 22 to 34 at time of Census survey |
| $\text{Dem}_{Senior}$ | Proportion of residents over 62 at time of Census survey |
| $\text{Dem}_{Jobs}$ | Proportion of Pop+Job that are jobs |
| $\boldsymbol{x}$ | Vector of explanatory variables |
| $t$ | Year $\in (0, ..., T)$ |
| $i$ | Route-segment $\in (0, ..., n)$ |

---

[2] Because LEHD did not have a non-Hispanic white tabulation, the distinction could not be made. For consistency, we used the same definition in ACS data.

## 4. Case studies

Four transit agencies from different regions of the United States were selected to explain the nation-wide bus ridership decline. These transit agencies are similar in key ways that make their comparison possible while being different enough to represent a wide cross-section of mid-sized to large transit agencies. They also align with the national trends; between 2012 and 2018, bus ridership declined by 13.0% nationally, 6.01 % at TriMet, 35.2 % at Miami-Dade Transit, 21.4% at Metro Transit, and 13.6% at MARTA. (Federal Transit Administration, 2019) These four agencies were selected because they had several years of reliable APC data at full or almost full coverage. While the research team reached out to 14 agencies, only those included in this study fulfilled the criteria of consistency and completeness.

Table 2 shows the time span of data used in this analysis and the summary statistics of the regression variables. None of the transit agencies had data available for 2012 through 2018. The analysis starts in 2013 in Miami, in 2014 in Atlanta and ends in 2017 in Portland, Miami, and Minneapolis. Because ridership is subject to seasonality, only one markup was used for each year. We selected the season that maximized the number of markups in the analysis instead of using the same season for each agency. The transit agencies in this case study vary widely with respect to the model variables. The average stop serves 50.9 passengers per weekday in Portland, but only 36.4 in Atlanta. Average frequency is more homogeneous across agencies, ranging between 37.5 trips per day in Miami and 43.6 in Atlanta. Atlanta also has the lowest average population and jobs surrounding bus stops with less than half of any other city. Portland is by far the whitest city with, on average, 80% of white residents surrounding bus stops. Miami is the least educated city with almost half of residents surrounding bus stops having no college education. The proportion of millennials and seniors is similar across the four cities. Jobs represent between 31%, in Miami, and 39%, in Portland, of trip generators close to bus stops.

Table 2: Years of Data Availability and Summary Statistics of the Regression Variables

| Variables | Portland Mean | Portland St.Dev | Miami Mean | Miami St.Dev | Minneapolis/St-Paul Mean | Minneapolis/St-Paul St.Dev | Atlanta Mean | Atlanta St.Dev |
|---|---|---|---|---|---|---|---|---|
| First Year | 2012 | | 2013 | | 2012 | | 2014 | |
| Last Year | 2017 | | 2017 | | 2017 | | 2018 | |
| Markup | Spring | | Winter | | Winter | | Summer | |
| Rid | 50.9 | 62.0 | 45.3 | 49.6 | 39.6 | 63.2 | 36.4 | 41.7 |
| Freq | 40.4 | 28.1 | 37.5 | 20.6 | 40.8 | 26.1 | 43.6 | 18.2 |
| Pop+Job | 1,451 | 2,004 | 1,439 | 1,573 | 1,681 | 2,435 | 710 | 915 |
| $\text{Dem}_{ZeroVehHH}$ | 0.13 | 0.11 | 0.13 | 0.10 | 0.15 | 0.10 | 0.16 | 0.11 |
| $\text{Dem}_{White}$ | 0.80 | 0.09 | 0.66 | 0.29 | 0.66 | 0.18 | 0.35 | 0.29 |
| $\text{Dem}_{HighSch}$ | 0.22 | 0.13 | 0.47 | 0.16 | 0.27 | 0.13 | 0.33 | 0.16 |
| $\text{Dem}_{Millennial}$ | 0.23 | 0.08 | 0.19 | 0.04 | 0.26 | 0.09 | 0.23 | 0.08 |
| $\text{Dem}_{Senior}$ | 0.17 | 0.05 | 0.19 | 0.06 | 0.15 | 0.05 | 0.15 | 0.06 |
| $\text{Dem}_{Jobs}$ | 0.39 | 0.25 | 0.31 | 0.24 | 0.38 | 0.25 | 0.35 | 0.25 |

## 5. Model

In this section, we first present the models evaluating which local socio-demographics are associated with change in bus ridership. These models are based on a framework developed by the authors to control for changes in frequency (Berrebi et al., 2020). We begin with a model to investigate the connection between neighborhood characteristics and ridership at one point in time. This cross-sectional analysis evaluates how the social, racial, and generational composition surrounding a route-segment relates to bus ridership when controlling for population, jobs, and frequency. Then, in order to model the ridership change, we present a model that captures the variation taking place within each individual route-segment over time. The Poisson fixed-effects model evaluates the relationship between neighborhood socio-demographics and the change in ridership over time while controlling for the change in frequency, population, and jobs.

### 5.1. Cross-Section

To model bus ridership at a point in time, it is important to take into consideration the particular shape of ridership data. Linear regression assumes that the data follows a Normal distribution, which is symmetric. The average daily boardings and alightings at a stop are necessarily non-negative and therefore skew to the right. A linear transformation that fits the count-nature of ridership data is the Poisson distribution. Poisson regression coefficients can be estimated with a Maximum Likelihood Estimation (MLE) technique. The Poisson regression is shown in Equations 1 and 2.

$$E[\text{Rid}|\boldsymbol{x}] = e^{(\beta_0 + \beta_1 * \log(\text{Freq}) + \beta_2 * \log(\text{Pop}+\text{Job}) + \beta_k * \text{Dem}_k)} \tag{1}$$

$$= e^{\beta_0} * \text{Freq}^{\beta_1} * (\text{Pop}+\text{Job})^{\beta_2} * e^{\beta_k * \text{Dem}_k} \tag{2}$$

The term $e^{\beta_0}$ gives the estimated ridership assuming that all explanatory variables are null. It is equivalent to the intercept in linear regression. Frequency is measured in total weekday vehicle-trips. Population and jobs are summed to form a trip generation variable. The Freq and (Pop+Job) variables are logged. Their coefficients, $\beta_1$ and $\beta_2$, are therefore measures of elasticity, i.e. what percentage change in ridership would result from a percentage change in explanatory variable. Note that the cross-sectional estimates of elasticity are only based on variation between individual route-segments. They cannot predict how changes in frequency at a particular route-segment would affect bus ridership.

Variables $k$ represent the proportion of population/jobs surrounding a route-segment from socio-demographic group $k$. When the proportion is equal to zero, the effect is null. When equal to one, i.e. the entire population, then Rid is multiplied by a factor $\beta_k$. When a parameter estimate, $\beta_k$, is positive, then the proportion of residents or jobs from group k is positively correlated with ridership. When negative, the opposite is true. For example, the demographic variable, $\text{Dem}_{jobs}$, assigns a coefficient to the proportion of trip generators that are jobs. It takes into account the different trip generation potential of job locations versus residential homes.

### 5.2. Fixed-Effects

The change in bus ridership is modeled with a Poisson fixed-effects model. The model includes individual-specific dummy variables, $\alpha_i$, to control for the variation taking place between route-segments and markups. The fixed-effects model captures the variation within each individual route-segment over time. Parameter estimates for the remaining explanatory variables therefore only explain the change in ridership. A linear time-trend, $\mu t$, captures the yearly ridership change that is not explained by other variables. As in the cross-sectional model, ridership data is assumed to follow a Poisson process. Equations 3 and 4 show the Poisson fixed-effects model with static socio-demographics.

$$E[\text{Rid}_{it}|\boldsymbol{x}_{it}] = e^{\left(\beta_1 * \log(\text{Freq}_{it}) + \beta_2 * \log(\text{Pop}_{it}+\text{Job}_{it}) + \beta_k * (\text{Dem}_k)_{it_0} * t + \alpha_i + \mu t\right)} \tag{3}$$

$$= \text{Freq}_{it}^{\beta_1} * (\text{Pop}_{it} + \text{Job}_{it})^{\beta_2} * e^{\beta_k * (\text{Dem}_k)_{it_0} * t} * e^{\alpha_i} * e^{\mu t} \tag{4}$$

In Equation 4, the $\beta_1$ and $\beta_2$ parameter estimates represent the elasticity of ridership to Freq and (Pop+Job). Unlike Equation 2, these elasticity estimates correspond to the ridership change resulting from frequency change at the individual route-segment level.

Static socio-demographic variables are interacted with time itself to explain the change in ridership (Wooldridge, 2002, §10.5). The $\beta_k$ coefficients can be interpreted as the change in their corresponding $\beta_k$ parameters of the cross-section model of Equation (1). They tell us whether the value of explanatory variables at time $t_0$ is correlated with the relative *change* in ridership over the years. In other words, these terms inform which socio-demographic characteristics are associated with ridership increase or decrease.

To control for the effect of demographic change on bus ridership, we simply add the time-varying socio-demographic term, $e^{\beta_l*(\text{Dem}_k)_{it}}$, to Equation 4 as follows:

$$E[\text{Rid}_{it}|\boldsymbol{x}_{it}] = \text{Freq}_{it}^{\beta_1} * (\text{Pop}_{it} + \text{Job}_{it})^{\beta_2} * e^{\beta_k*(\text{Dem}_k)_{it_0}*t} * e^{\beta_l*(\text{Dem}_k)_{it}} * e^{\alpha_i} * e^{\mu t} \tag{5}$$

In Equation 5, $\beta_l$ is the parameter estimate corresponding to the $k^{th}$ socio-demographic variable. This term captures the impact of shifting neighborhood characteristics on the change in bus ridership over time within each route-segment.

## 6. Bus Ridership as a Cross-Section

We begin by modeling bus ridership at a fixed point in time. Table 3 shows the cross-section model results. Ridership in the first markup (Spring 2012 for Portland, Winter 2013 for Miami, Winter 2012 for Minneapolis/St-Paul, and Summer 2014 for Atlanta) is explained with the static frequency, population, jobs, and socio-demographics. We use the McFadden pseudo-$R^2$ to evaluate the fit. The high pseudo-$R^2$'s indicate that the models' estimated likelihoods are high compared to models without any predictors.

While the concentration of millennial residents can be positively or negatively correlated, the presence of seniors is associated with lower bus ridership in all four cities. In Miami, the parameter estimate for millennials is significantly negative, whereas in Minneapolis/St-Paul it is significantly positive. This result may be a reflection of the different types of places where millennials live in both cities. Regarding seniors, the parameter estimates are consistent in sign and significance in all four cities. Neighborhoods with high concentrations of residents above 62 have lower ridership when controlling for frequency, population, and jobs.

In all four cities, the share of jobs in total trip generators (populations + jobs) is positive, which indicates that each job contributes more to ridership than each individual resident. This may be because areas with high job to total trip generator ratio tend to be highly concentrated, and therefore transit-supportive.

Finally, race, education, and vehicle ownership all correlate with bus usage. In Atlanta and Minneapolis/St-Paul, the proportion of zero vehicle households is positively associated with ridership. In Miami, the proportion of white residents is negatively correlated with bus ridership. In all four agencies, the proportion of residents whose maximum level of education is high-school correlates significantly and positively with ridership. These results are consistent with the literature (Giuliano, 2005); particularly with Dill et al. (2013), who modeled ridership at the stop-level in Portland as a cross-section analysis.

Table 3: Cross-Section Models of Bus Ridership at One Point in Time (First Markup)

|  | Response Variable: | Rid | | |
| --- | --- | --- | --- | --- |
|  | Portland | Miami | Minneapolis/St-Paul | Atlanta |
| log(Freq) | 1.35 (0.04)*** | 1.19 (0.03)*** | 1.02 (0.02)*** | 1.20 (0.08)*** |
| log(Pop + Job) | 0.41 (0.06)*** | 0.65 (0.04)*** | 0.43 (0.05)*** | 0.20 (0.06)** |
| $\text{Dem}_{ZeroVehHH}$ | 0.32 (0.31) | 0.34 (0.18)· | 0.80 (0.35)* | 1.36 (0.40)*** |
| $\text{Dem}_{White}$ | −0.44 (0.26)· | −0.24 (0.08)** | 0.15 (0.24) | −0.05 (0.17) |
| $\text{Dem}_{HighSch}$ | 0.58 (0.19)** | 0.41 (0.12)*** | 1.37 (0.35)*** | 0.59 (0.30)* |
| $\text{Dem}_{Millennial}$ | −0.15 (0.35) | −3.18 (0.48)*** | 1.52 (0.30)*** | 0.92 (0.48)· |
| $\text{Dem}_{Senior}$ | −1.34 (0.50)** | −1.60 (0.39)*** | −2.07 (0.55)*** | −1.67 (0.65)* |
| $\text{Dem}_{Jobs}$ | 0.46 (0.12)*** | 0.16 (0.09)· | 0.24 (0.15) | 0.61 (0.17)*** |
| Intercept | −3.87 (0.49)*** | −4.44 (0.24)*** | −4.15 (0.45)*** | −2.87 (0.49)*** |
| Pseudo $R^2$ | 0.83 | 0.75 | 0.81 | 0.45 |
| Deviance | 8829.85 | 11730.70 | 16961.48 | 12979.48 |
| Num. obs. | 874 | 1165 | 1862 | 718 |

***$p < 0.001$; **$p < 0.01$; *$p < 0.05$; ·$p < 0.1$

In Figure 1, the relationship between bus ridership per trip per capita (in Pop + Job) and education is plotted. The regression line in red is surrounded by a gray confidence band representing one standard error. There is a clear positive relationship in all four cities between ridership and the proportion of residents whose maximum level of education is high school.

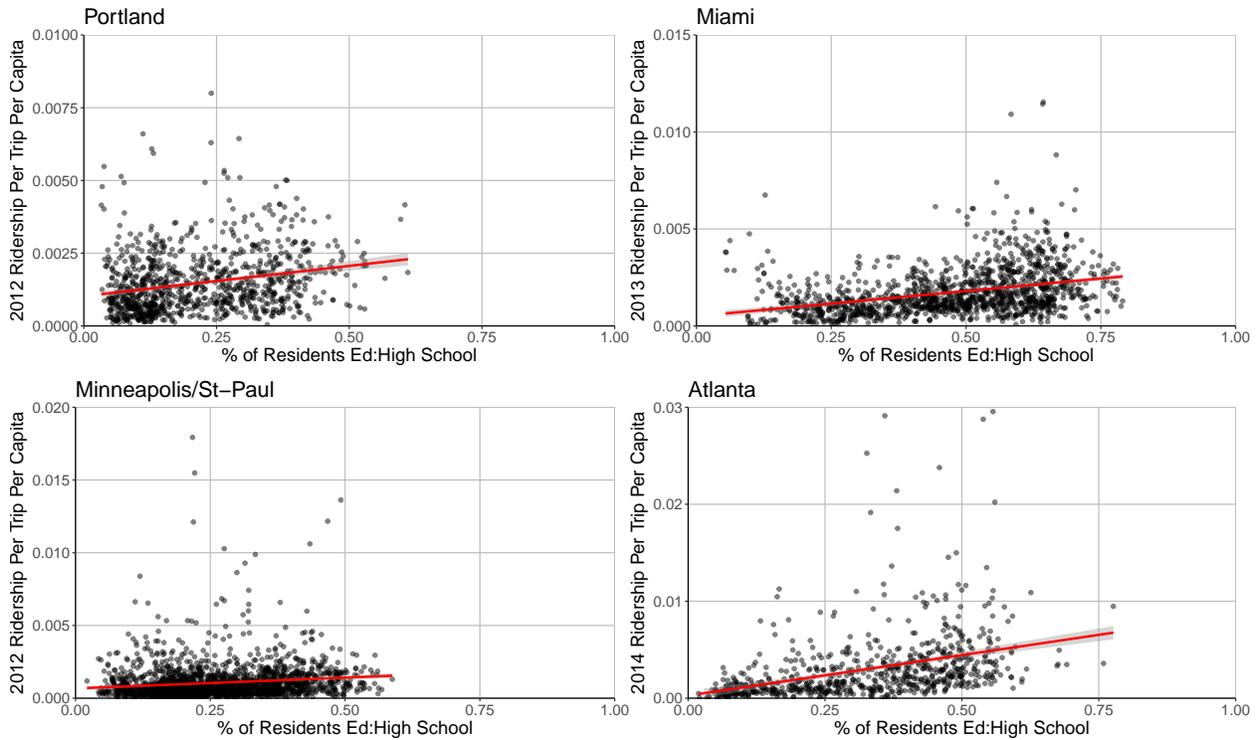

Figure 1: First Markup Bus Ridership Per Trip Per Capita Versus Proportion of Residents Whose Maximum Level of Education is High School

## 7. Bus Ridership as a Panel with Static Socio-Demographics

We now model the change in bus ridership over time as a function of frequency, population, and job change, and of static socio-demographics. Table 4 shows the fixed-effects model results. Unfortunately, the Poisson fixed-effects model does not have an equivalent to $R^2$. The variable $t$ represents the ridership change across route-segments, which is otherwise unaccounted for in the model. It is only significant in Atlanta and Minneapolis/St-Paul, where it can be interpreted as a systematic effect pulling ridership up or down. In the other two agencies, the time-intercept is not significant, indicating that the model explains the overall decline.

Most variables differ in sign and significance among agencies. The effects of population and job change over time is only significant in Atlanta and Minneapolis/St-Paul. Even in these cities, the coefficient is relatively low. This is consistent with Ederer et al. (2019), which found a significant effect of population change on rail ridership at the regional level but no impact on bus. The advantage of jobs versus population in generating ridership strengthened in Miami and Atlanta and diminished in Minneapolis/St-Paul. There is no consistent change in age-related bus ridership across cities. The presence of both seniors and millennials correlate with ridership gain in Portland and with ridership loss in Miami. Neighborhoods were more likely to increase in ridership if they had high concentrations of seniors in Minneapolis/St-Paul and millennials in Atlanta.

The proportion of white residents is the only socio-demographic factor that consistently and significantly affects ridership across all four agencies. In every case study, white neighborhoods were more likely to decline in ridership when controlling for frequency, population, and job change. In Miami and Atlanta, the proportion of residents without a college education is associated with ridership gains and the proportion of residents without access to a car was associated with ridership loss. These results indicate that, in these two cities, the recent decline in ridership is happening in white, educated, and carless neighborhoods. In Minneapolis/St-Paul, the proportion of residents who did not attend college is correlated with ridership decline.

Table 4: Poisson Fixed-Effects Models of Bus Ridership Change with Static Socio-Demographics

|  | Response Variable: | $\text{Rid}_{it}$ | | |
| --- | --- | --- | --- | --- |
|  | Portland | Miami | Minneapolis/St-Paul | Atlanta |
| $\log(\text{Freq}_{it})$ | 0.71 (0.04)*** | 0.82 (0.04)*** | 0.78 (0.03)*** | 0.69 (0.01)*** |
| $\log(\text{Pop}_{it} + \text{Job}_{it})$ | −0.02 (0.04) | −0.02 (0.03) | 0.08 (0.03)* | 0.05 (0.01)*** |
| $(\text{Dem}_{ZeroVehHH})_{it_0} * t$ | 0.01 (0.06) | −0.26 (0.07)*** | −0.11 (0.07) | −0.39 (0.04)*** |
| $(\text{Dem}_{White})_{it_0} * t$ | −0.18 (0.07)* | −0.12 (0.03)*** | −0.23 (0.06)*** | −0.12 (0.02)*** |
| $(\text{Dem}_{HighSch})_{it_0} * t$ | 0.02 (0.05) | 0.22 (0.05)*** | −0.23 (0.08)** | 0.27 (0.03)*** |
| $(\text{Dem}_{Millennial})_{it_0} * t$ | 0.38 (0.10)*** | −0.55 (0.20)** | −0.06 (0.08) | 0.25 (0.05)*** |
| $(\text{Dem}_{Senior})_{it_0} * t$ | 0.37 (0.14)** | −0.42 (0.17)* | 0.49 (0.13)*** | −0.08 (0.06) |
| $(\text{Dem}_{Jobs})_{it_0} * t$ | −0.02 (0.03) | 0.13 (0.03)*** | −0.16 (0.02)*** | 0.06 (0.01)*** |
| t | −0.02 (0.02) | −0.01 (0.02) | 0.04 (0.02)* | −0.06 (0.01)*** |
| Log-Likelihood | −10117.52 | −10303.07 | −17249.19 | −19333.60 |
| Num. obs. | 4884 | 5264 | 10929 | 3453 |
| n | 874 | 1165 | 1862 | 718 |
| T | 6 | 5 | 6 | 5 |

***$p < 0.001$; **$p < 0.01$; *$p < 0.05$; ·$p < 0.1$

In order to investigate the main trend found in Table 4, Figure 2 shows the log-relative change in bus ridership versus the proportion of white residents within a 1/4 mile radius surrounding route-segments. The vertical axis shows the log of ridership in the last markup divided by the first markup, which is how the Poisson model considers change in the response variable. The opacity of each point is weighted by its first-year productivity in ridership per trip. The red line shows a simple regression trend with every route-segment weighted the same. The green line shows the regression trend weighted by first-year productivity [3]. Lighter points, which represent low-productivity route-segments, have considerably more variation. Since these segments have fewer boardings and alightings per trip to begin with, a few more or less passengers can overwhelm the log-relative change in ridership.

There is a consistent negative relationship between the proportion of white residents and the change in ridership, but it does not entirely explain the overall bus ridership decline. In all four agencies, whiter neighborhoods lost more ridership, relatively, even when not controlling for population, jobs, and all other socio-demographic variables. This trend is strongest in Portland and Atlanta; less so in Miami and Minneapolis/St-Paul. In Minneapolis/St-Paul, the weighted regression line is steeper, which means that $\text{Dem}_{White}$ has a greater effect in high-productivity segments.

---

[3] Note that all model results in this paper are unweighted.

The trend, however, does not account for all of the ridership decline. Route-segments surrounded by low proportions of white residents also lost ridership but not as much as homogeneously white neighborhoods. In Miami, Minneapolis/St-Paul, and Atlanta, places with just 50% of white residents were still expected to drop in productivity. Portland is the exception but almost all route-segments are majority-white.

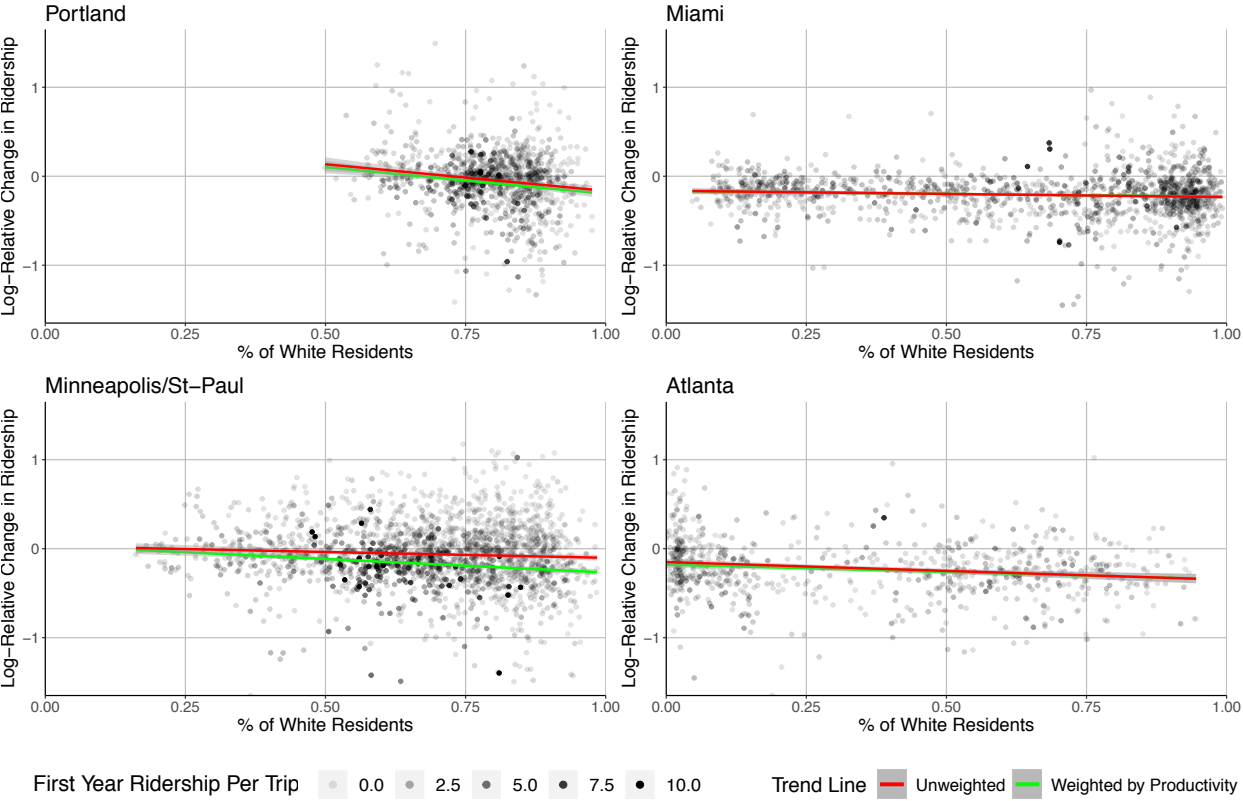

Figure 2: Log-Relative Change in Bus Ridership Versus Proportion of White Residents

## 8. Bus Ridership as a Panel with Shifting Demographics

The previous section established the connection between local demographics and bus ridership change assuming that neighborhoods' compositions have remained constant since 2012. These trends, however, could be confounded with demographic shifts resulting from intra-city migration. In this section, we control for shifts in local demographics. As discussed in the literature review, gentrification has typically been concentrated in places where transit is most available. For the purpose of this study, we define gentrification as changes in local socio-demographics irrespective of previous characteristics. The assumption is that a change in $\text{Dem}_k$ from 0 to 10% would have the same effect as a change from 90% to 100%

### 8.1. Assessing the Magnitude of Demographic Shifts Near Bus Service

We begin by assessing how demographics may have shifted near bus service. Race is used as the variable of interest because, unlike education and income, it remains constant over the course of one's life. Figure 3 shows a histogram of change in proportion of white resident workers between 2011 and 2015. Note that all route-segments are weighted evenly regardless of population and service. In Portland, the distribution is narrow and centered with 89.8% of route-segments in 2015 within three percentage points of 2011 $\text{Dem}_{White}$. In Minneapolis/St-Paul, the change in proportion of white residents is more widely distributed around zero with only 69.0% of route-segments within three percentage points. In Atlanta, the distribution is close to symmetrical around its positive median, which indicates that the proportion of white residents increased in most segments. Finally, in Miami, the distribution is centered near zero ($\text{Dem}_{White}$ increased at 50.4% of segments) but skews to the right with 6.78% of segments where the proportion of white residents increased by more than 10 percentage points.

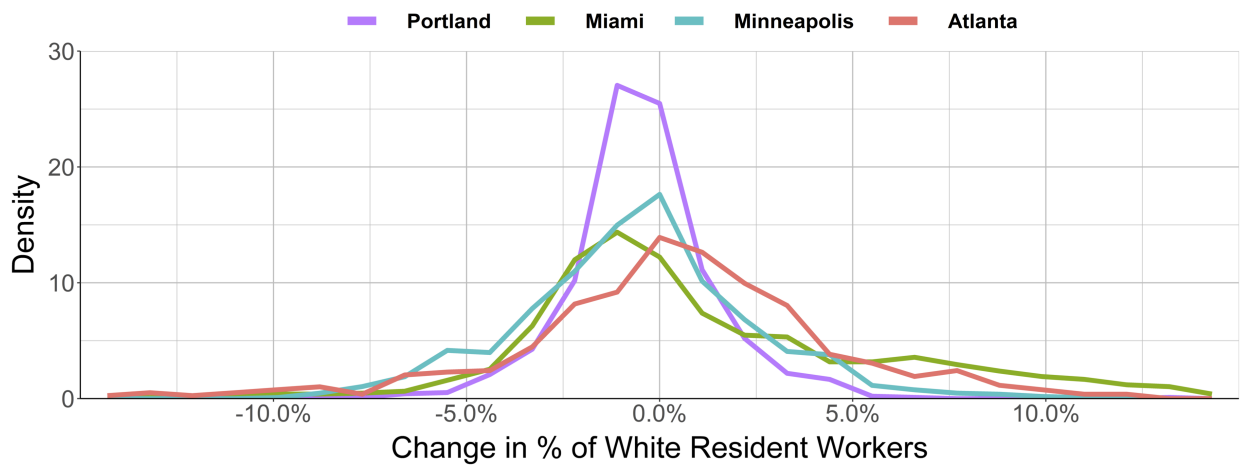

Figure 3: Histogram of Change in $\text{Dem}_{White}$ Between 2011 and 2015

In order to make the connection between the national bus ridership decline and demographic shift, the magnitude of the phenomenon must be quantified. Table 5 shows the network-level change in proportion of white residents. The first row contains the change in percent of white residents over the entire service area. People moving from transit-oriented neighborhoods to low-density suburbs may still have access to a bus stop but would be far less likely to use it. To take into consideration the temporal dimension of transit accessibility, the second row of Table 5 shows the change in percent of white residents over the entire network weighted by the service frequency at nearby segments. In other words, someone living near a high frequency route is counted more heavily than someone living in a low-service area.

In Portland and Minneapolis, the proportion of white residents has declined overall, while it increased in Miami and Atlanta. In Portland and Minneapolis, the weighted change is greater than the non-weighted change, which indicates that the decrease in $\text{Dem}_{White}$ is concentrated in low-service areas. In Atlanta, the overall increase in proportion of white residents was offset by a decrease in the most concentrated areas of service. When weighting by frequency, the proportion of white residents decreased overall. Miami is the only place where the places having access to bus service on both a spatial and spatio-temporal level have become whiter.

Table 5: Network-Level Change in Proportion of White Residents

|  | Portland | Miami | Minneapolis/St-Paul | Atlanta |
|---|---|---|---|---|
| Change in % of $\text{Dem}_{White}$ | -0.44 | 0.35 | -0.70 | 0.39 |
| Change in % of $\text{Dem}_{White}$ Weighted by Freq | -0.36 | 0.23 | -0.48 | -0.17 |

Overall, demographic shifts in transit agencies' service areas are subtle. The proportion of white residents is only increasing in Miami and Atlanta. Even there, the network-wide change is less than 0.5 percentage points. The growth in proportion of white population is concentrated in areas with little bus service. These results indicate that intra-urban migration taking place between 2011 and 2015 was limited. Note, however, that the LEHD data is based on employment. Since the employment rate increased between 2011 and 2015, especially for non-white workers, the demographic shifts are likely to be underestimated in LEHD data (Bureau of Labor Statistics, 2018). They might, however, help control for some of the ridership change occurring on a hyper-local level in places where the groups most closely associated with bus ridership are declining in proportion.

*8.2. Modeling the Impact of Demographic Shifts on Bus Ridership*

The Poisson fixed-effects models of ridership change with static demographics and demographic change are shown in Table 6. The variables $(\text{Dem}_{White})_{it}$ and $(\text{Dem}_{HighSch})_{it}$ represent the changing proportion of white and high-school educated residents. Other demographic variables could not be included because the LEHD does not provide information on vehicle ownership, and the age categories are not consistent with ACS.

The time-varying frequency, population, and job, as well as static demographic parameter estimates in Table 6, are consistent with the results from Table 4. In particular, the proportion of white residents remains a significant predictor of ridership decline in all four cities. In addition, the proportion of residents with a college education and without access to cars are still linked to ridership decline in Miami and Atlanta. In Minneapolis, the proportion of resident without a college education is still correlated with ridership decline.

The results differ between cities. In Portland, demographic shifts between 2011 and 2015 have no significant effect on ridership. In Miami, the increasing proportion of white residents is slightly correlated with ridership decline on a local level while education is insignificant. There, the increase in overall proportion of white residents, including in areas of highest service concentration, could therefore be a factor contributing to the ridership decline. In Minneapolis/St-Paul, the growth in proportion of high-school educated residents is correlated with a decline in ridership. In Atlanta, increases in the proportions of high-school educated and white residents are correlated with ridership increase. The estimated effects of $(\text{Dem}_{White})_{it}$ and $(\text{Dem}_{HighSch})_{it}$ are not an artifact of their co-linearity. Appendix A shows that, even when considered separately, their sign and significance persist. The magnitude of demographic shifts captured by LEHD data, however, indicates that their overall impact may be limited. More precise data over a longer period of time would be needed to fully uncover the impact of demographic shifts on ridership.

Table 6: Poisson Fixed-Effects Models of Bus Ridership Change with Static Demographics and Demographic Change

|  | Response Variable: | $\text{Rid}_{it}$ |  |  |
| --- | --- | --- | --- | --- |
|  | Portland | Miami | Minneapolis/St-Paul | Atlanta |
| $\log(\text{Freq}_{it})$ | 0.71 (0.04)*** | 0.81 (0.04)*** | 0.78 (0.03)*** | 0.69 (0.01)*** |
| $\log(\text{Pop}_{it} + \text{Job}_{it})$ | −0.02 (0.04) | −0.02 (0.03) | 0.07 (0.03)* | 0.05 (0.01)*** |
| $(\text{Dem}_{ZeroVehHH})_{it_0} * t$ | 0.02 (0.06) | −0.26 (0.07)*** | −0.09 (0.08) | −0.39 (0.04)*** |
| $(\text{Dem}_{White})_{it_0} * t$ | −0.19 (0.07)** | −0.14 (0.03)*** | −0.21 (0.06)*** | −0.10 (0.02)*** |
| $(\text{Dem}_{HighSch})_{it_0} * t$ | 0.03 (0.05) | 0.22 (0.06)*** | −0.21 (0.08)* | 0.26 (0.03)*** |
| $(\text{Dem}_{Millennial})_{it_0} * t$ | 0.38 (0.10)*** | −0.57 (0.21)** | −0.09 (0.08) | 0.25 (0.05)*** |
| $(\text{Dem}_{Senior})_{it_0} * t$ | 0.37 (0.14)** | −0.45 (0.17)** | 0.43 (0.13)** | −0.06 (0.06) |
| $(\text{Dem}_{Jobs})_{it_0} * t$ | −0.03 (0.03) | 0.14 (0.03)*** | −0.16 (0.02)*** | 0.05 (0.01)*** |
| $(\text{Dem}_{White})_{it}$ | −0.20 (0.27) | −0.28 (0.13)* | −0.26 (0.18) | 0.45 (0.08)*** |
| $(\text{Dem}_{HighSch})_{it}$ | 0.22 (0.26) | −0.03 (0.25) | −1.21 (0.37)*** | 0.31 (0.10)** |
| t | −0.02 (0.02) | 0.00 (0.02) | 0.05 (0.02)** | −0.07 (0.01)*** |
| Log-Likelihood | −10116.86 | −10300.74 | −17242.70 | −19313.24 |
| Num. obs. | 4884 | 5264 | 10929 | 3453 |
| n | 874 | 1165 | 1862 | 718 |
| T | 6 | 5 | 6 | 5 |

***$p < 0.001$; **$p < 0.01$; *$p < 0.05$; ·$p < 0.1$

## 9. Conclusion

### 9.1. The Local Characteristics of Bus Ridership Decline

In all four cities, some of the socio-demographic groups associated with the least bus ridership at a point in time are responsible for the most ridership decline over time. This study confirms that the socio-demographic factors associated with bus ridership in the literature correlate with cross-sectional ridership at the route-segment level. While ridership declined across neighborhood characteristics between 2012 and 2018, in all four cities, ridership declined the most in white neighborhoods when controlling for the change in frequency, population, and jobs. Even on its own, i.e. without controlling for other variables, there is a clear negative relationship between the proportion of white residents and the relative change in bus ridership. In addition, ridership change correlates positively with the proportion of high-school educated residents and negatively with the carlessness rate in Miami and Atlanta. In Minneapolis/St-Paul, the proportion of residents who are not college-educated is linked to bus ridership decline. The sign and significance of these results remain consistent even when controlling for demographic shifts happening over time.

While these findings do not explain **why** bus ridership is declining, they provide an important clue: some of the underlying ridership decline's underlying causes may be affecting travel behaviors in all types of neighborhoods, but especially in places where white people live.

- One trend that has affected travel behaviors of white and educated employees is the upsurge of telecommuting (Walls et al., 2007). Between 2012 and 2016, the share of US employees working remotely at least some of the time rose from 39% to 43% and they spent more time doing so (Gallup, 2017).

- Another possible explanation is the rise of ridehailing. Several studies report that people who are white, college-educated, and have low-vehicle access are more likely to use ridehailing (Rayle et al., 2016; Dias et al., 2017; Circella et al., 2018; Sikder, 2019; Henao and Marshall, 2019). Ridehailing more than doubled the size of the overall for-hire services sector between 2012 and 2017, when they were expected to surpass overall bus ridership in the United States (Schaller, 2018).

Trends in telecommuting and ridehailing therefore coincide with the ridership decline across time and neighborhood socio-demographics, especially in Miami and Atlanta. In Minneapolis, where the proportion of local residents who are white is correlated with ridership loss and the proportion of college-educated residents is associated with ridership gain, these phenomena are less likely to be the main driving factors.

The analysis of demographic changes does not show a consistent effect pulling bus ridership downward. In the LEHD dataset, the proportion of white residents only increased overall in Miami and Atlanta. Even there, the growth was moderate and concentrated near low-frequency routes. When considering how the time-varying proportion of white and high-school educated residents may affect ridership change, we do not find a consistent effect between the two metrics in any of the cities. Only in Miami could the increasing proportion of white residents near transit help explain the decline in bus ridership. In other cities, the effect is null, positive, or inconsistent. This part of the analysis, however, is based on a crude dataset; LEHD only includes residents who are employed and counts people with multiple jobs multiple times. In addition, the limited time-span, from 2011 to 2015, could explain why the impact of gentrification on ridership found by Mills and Steele (2017) in Portland was not significant in our analysis. With access to demographic data on a wider time-frame, it is possible that more precise ridership patterns would emerge. In the future, when reliable APC data become available for ten consecutive year, two non-overlapping ACS five-year estimates could be used to explore in greater detail the effect of gentrification on ridership.

*9.2. Future Research*

One of the key missing pieces in this analysis is the rail network. Since the early 1990s transit agencies started prioritizing rail service over buses in an effort to capture potential riders who have access to private cars. This effort culminated in 2017 when rail ridership surpassed bus ridership for the first time[4] since the dismantlement of the American streetcar system (Dickens, 2018). Rail ridership directly affects bus ridership as riders transfer from one mode to the other. The few connection points between bus and rail typically have disproportionately high ridership, which is not related to the surrounding population, jobs, and socio-demographics. These stops, however, have little leverage on the Poisson Fixed-Effects model, which measures the relative change in ridership over time within each route-segments. Nonetheless, future research should verify whether rail stations located in mostly white neighborhoods have increased in ridership as a result of mode substitution or if, on the contrary, the same trends exhibited in the bus network apply to rail. In either case, the findings may have profound policy ramifications.

## 10. Acknowledgments

The authors wish to extend heartfelt gratitude for staff members at TriMet, Miami-Dade Transit, Metro Transit, and MARTA for providing us with the data and helping us understand their transit systems. In particular, we thank Nathan Banks at Trimet, Eric Lind, Janet Hopper, and John Levin, at Metro Transit, Esther Frometa-Spring at Miami-Dade Transit, and Nazma Akhter, Lekha Mukherjee, Ivelisse Matos, and Robert Goodwin at MARTA. Finally, we thank Rashad Strickland from Clever Devices and Thomas Kowalsky from Urban Transportation Associates for their help navigating the data and understanding the technology.

---

[4] albeit only for a year

# 11. Appendix A

|  | Response Variable: | $\text{Rid}_{it}$ | | |
|---|---|---|---|---|
|  | Portland | Miami | Minneapolis/St-Paul | Atlanta |
| $\log(\text{Freq}_{it})$ | 0.71 (0.04)*** | 0.81 (0.04)*** | 0.78 (0.03)*** | 0.69 (0.01)*** |
| $\log(\text{Pop}_{it} + \text{Job}_{it})$ | −0.02 (0.04) | −0.02 (0.03) | 0.08 (0.03)* | 0.05 (0.01)*** |
| $(\text{Dem}_{ZeroVehHH})_{it_0} * t$ | 0.01 (0.06) | −0.26 (0.07)*** | −0.08 (0.08) | −0.39 (0.04)*** |
| $(\text{Dem}_{White})_{it_0} * t$ | −0.19 (0.07)** | −0.14 (0.03)*** | −0.23 (0.06)*** | −0.11 (0.02)*** |
| $(\text{Dem}_{HighSch})_{it_0} * t$ | 0.02 (0.05) | 0.22 (0.05)*** | −0.24 (0.08)** | 0.26 (0.03)*** |
| $(\text{Dem}_{Millennial})_{it_0} * t$ | 0.39 (0.10)*** | −0.56 (0.20)** | −0.07 (0.08) | 0.24 (0.05)*** |
| $(\text{Dem}_{Senior})_{it_0} * t$ | 0.36 (0.14)** | −0.44 (0.17)** | 0.43 (0.13)** | −0.05 (0.06) |
| $(\text{Dem}_{Jobs})_{it_0} * t$ | −0.03 (0.03) | 0.14 (0.03)*** | −0.16 (0.02)*** | 0.06 (0.01)*** |
| $(\text{Dem}_{White})_{it}$ | −0.21 (0.27) | −0.28 (0.13)* | −0.27 (0.18) | 0.44 (0.08)*** |
| t | −0.02 (0.02) | 0.00 (0.02) | 0.05 (0.02)** | −0.07 (0.01)*** |
| Log-Likelihood | −10117.22 | −10300.75 | −17248.12 | −19318.21 |
| Num. obs. | 4884 | 5264 | 10929 | 3453 |
| n | 874 | 1165 | 1862 | 718 |
| T | 6 | 5 | 6 | 5 |

***$p < 0.001$; **$p < 0.01$; *$p < 0.05$; ˙$p < 0.1$

Table 7: Poisson Fixed-Effects Models of Ridership Change with Static Demgraphics and $\text{Dem}_{White}$ Change

|  | Response Variable: | $\text{Rid}_{it}$ | | |
|---|---|---|---|---|
|  | Portland | Miami | Minneapolis/St-Paul | Atlanta |
| $\log(\text{Freq}_{it})$ | 0.71 (0.04)*** | 0.82 (0.04)*** | 0.78 (0.03)*** | 0.69 (0.01)*** |
| $\log(\text{Pop}_{it} + \text{Job}_{it})$ | −0.02 (0.04) | −0.02 (0.03) | 0.07 (0.03)* | 0.05 (0.01)*** |
| $(\text{Dem}_{ZeroVehHH})_{it_0} * t$ | 0.02 (0.06) | −0.26 (0.07)*** | −0.12 (0.07) | −0.39 (0.04)*** |
| $(\text{Dem}_{White})_{it_0} * t$ | −0.18 (0.07)* | −0.12 (0.03)*** | −0.20 (0.06)*** | −0.12 (0.02)*** |
| $(\text{Dem}_{HighSch})_{it_0} * t$ | 0.03 (0.05) | 0.21 (0.06)*** | −0.20 (0.08)* | 0.27 (0.03)*** |
| $(\text{Dem}_{Millennial})_{it_0} * t$ | 0.38 (0.10)*** | −0.56 (0.21)** | −0.08 (0.08) | 0.26 (0.05)*** |
| $(\text{Dem}_{Senior})_{it_0} * t$ | 0.37 (0.14)** | −0.43 (0.17)* | 0.49 (0.13)*** | −0.09 (0.06) |
| $(\text{Dem}_{Jobs})_{it_0} * t$ | −0.03 (0.03) | 0.13 (0.03)*** | −0.17 (0.02)*** | 0.05 (0.01)*** |
| $(\text{Dem}_{HighSch})_{it}$ | 0.23 (0.26) | −0.11 (0.25) | −1.22 (0.37)*** | 0.29 (0.10)** |
| t | −0.02 (0.02) | −0.00 (0.02) | 0.04 (0.02)** | −0.07 (0.01)*** |
| Log-Likelihood | −10117.12 | −10302.98 | −17243.68 | −19329.11 |
| Num. obs. | 4884 | 5264 | 10929 | 3453 |
| n | 874 | 1165 | 1862 | 718 |
| T | 6 | 5 | 6 | 5 |

***$p < 0.001$; **$p < 0.01$; *$p < 0.05$; ˙$p < 0.1$

Table 8: Poisson Fixed-Effects Models of Ridership Change with Static Demgraphics and $\text{Dem}_{HighSch}$ Change

**Bibliography**

Andersson, F., Garcia-Perez, M., Haltiwanger, J., McCue, K., Sanders, S., 2014. Workplace concentration of immigrants. Demography 51, 2281–2306.

Berrebi, S., Gibbs, T., Joshi, S., Watkins, K.E., 2020. On ridership and frequency. arXiv preprint: 2002.02493 .

Boisjoly, G., Grisé, E., Maguire, M., Veillette, M.P., Deboosere, R., Berrebi, E., El-Geneidy, A., 2018. Invest in the ride: A 14 year longitudinal analysis of the determinants of public transport ridership in 25 north american cities. Transportation Research Part A 116, 434–445.

Brown, A.E., Blumenber, E., Taylor, B.D., Ralph, K., Voulgaris, C.T., 2016. A taste for transit? analyzing public transit use trends among youth .

Brown-Saracino, J., 2017. Explicating divided approaches to gentrification and growing income inequality. Annual Review of Sociology 43, 515–539.

Bureau of Labor Statistics, 2018. Labor Force Characteristics by Race and Ethnicity, 2017. URL: https://www.bls.gov/opub/reports/race-and-ethnicity/2017/home.htm.

Chen, C., Varley, D., Chen, J., 2011. What affects transit ridership? a dynamic analysis involving multiple factors, lags and asymmetric behaviour. Urban Studies 48, 1893–1908.

Circella, G., Alemi, F., Tiedeman, K., Handy, S., Mokhtarian, P., 2018. The adoption of shared mobility in california and its relationship with other components of travel behavior .

Coogan, M., Spitz, G., Adler, T., McGuckin, N., Kuzmyak, R., Karash, K., 2018. Understanding changes in demographics, preferences, and markets for public transportation. Technical Report.

Dias, F.F., Lavieri, P.S., Garikapati, V.M., Astroza, S., Pendyala, R.M., Bhat, C.R., 2017. A behavioral choice model of the use of car-sharing and ride-sourcing services. Transportation 44, 1307–1323.

Dickens, M., 2018. Ridership by mode and quarter 1990 present. American Public Transit Association.

Dickens, M., MacPherson, H.C., 2019. APTA 2019 Public Transportation FactBook, Appendix A .

Dill, J., Schlossberg, M., Ma, L., Meyer, C., 2013. Predicting transit ridership at the stop level: The role of service and urban form, in: 92nd annual meeting of the Transportation Research Board, Washington, DC.

Driscoll, R.A., Lehmann, K.R., Polzin, S., Godfrey, J., 2018. The effect of demographic changes on transit ridership trends. Transportation Research Record , 0361198118777605.

Ederer, D., Berrebi, S., Diffee, C., Gibbs, T., Watkins, K.E., 2019. Comparing transit agency peer groups using cluster analysis. Transportation Research Record , 0361198119852074.

Ellen, I.G., O'Regan, K.M., 2011. How low income neighborhoods change: Entry, exit, and enhancement. Regional Science and Urban Economics 41, 89–97.

Federal Highway Administration, 2019. Technical Report. URL: https://www.fhwa.dot.gov/policyinformation/travel_monitoring/19augtvt/19augtvt.pdf.

Federal Transit Administration, 2019. [national transit database]. https://www.transit.dot.gov/ntd.

Freeman, L., 2005. Displacement or succession? residential mobility in gentrifying neighborhoods. Urban Affairs Review 40, 463–491.

Freeman, L., Cai, T., 2015. White entry into black neighborhoods: Advent of a new era? The ANNALS of the American Academy of Political and Social Science 660, 302–318.

Frei, C., Mahmassani, H.S., 2013. Riding more frequently: Estimating disaggregate ridership elasticity for a large urban bus transit network. Transportation Research Record 2350, 65–71.

Frey, W.H., 2018. US population disperses to suburbs, exurbs, rural areas, and of the country metros. The Avenue March 26. Brookings.

Gallup, 2017. State of the american workplace. Retrieved from Washington, DC: http://www.gallup.com .

Giuliano, G., 2005. Low income, public transit, and mobility. Transportation Research Record 1927, 63–70.

Glaeser, E.L., Kahn, M.E., Rappaport, J., 2008. Why do the poor live in cities? the role of public transportation. Journal of urban Economics 63, 1–24.

Glass, R., 1964. Aspects of change. The gentrification debates: A reader , 19–30.

Graehler Jr, M., Mucci, R.A., Erhardt, G.D., 2019. Understanding the Recent Transit Ridership Decline in Major US Cities: Service Cuts or Emerging Modes? Technical Report.

Graham, M.R., Kutzbach, M.J., McKenzie, B., et al., 2014. Design comparison of LODES and ACS commuting data products. Technical Report.

Grengs, J., 2001. Does public transit counteract the segregation of carless households? measuring spatial patterns of accessibility. Transportation Research Record 1753, 3–10.

Grimsrud, M., El-Geneidy, A., 2013. Driving transit retention to renaissance: trends in montreal commute public transport mode share and factors by age group and birth cohort. Public Transport 5, 219–241.

Grimsrud, M., El-Geneidy, A., 2014. Transit to eternal youth: lifecycle and generational trends in greater montreal public transport mode share. Transportation 41, 1–19.

Guerrieri, V., Hartley, D., Hurst, E., 2013. Endogenous gentrification and housing price dynamics. Journal of Public Economics 100, 45–60.

Hall, J.D., Palsson, C., Price, J., 2018. Is uber a substitute or complement for public transit? Journal of Urban Economics 108, 36–50.

Henao, A., Marshall, W.E., 2019. The impact of ride-hailing on vehicle miles traveled. Transportation 46, 2173–2194.

Higashide, S., Buchanan, M., 2019. Whos on board 2019 how to win back americas transit riders. Transit Center .

Hwang, J., Lin, J., 2016. What have we learned about the causes of recent gentrification? Cityscape 18, 9–26.

Hyra, D., 2016. Commentary: Causes and consequences of gentrification and the future of equitable development policy. Cityscape 18, 169–178.

Ihrke, D.K., Faber, C.S., Koerber, W.K., 2011. Geographical mobility: 2008 to 2009. US Department of Commerce, Economics and Statistics Administration, US .

Kain, J.F., Liu, Z., 1999. Secrets of success: assessing the large increases in transit ridership achieved by houston and san diego transit providers. Transportation Research Part A: Policy and Practice 33, 601–624.

Kittelson Associate, 2013. Transit capacity and quality of service manual.

Kneebone, E., 2017. The changing geography of us poverty. Brookings Institutions, February .

Kneebone, E., Holmes, N., 2014. New census data show few metro areas made progress against poverty in 2013. Metropolitan Opportunity Series report, Brookings Institution, Washington, DC .

Levinson, D.M., 1998. Accessibility and the journey to work. Journal of Transport Geography 6, 11–21.

Lu, X., Pas, E.I., 1999. Socio-demographics, activity participation and travel behavior. Transportation Research part A: policy and practice 33, 1–18.

Lyons, T.J., Tian, G., Daly, K., 2017. Understanding transit ridership in Utah: Improving methodologies, developing an accessible estimation technique, and establishing new relationships. Technical Report.

Maciag, M., 2015. Gentrification in america report. Governing the States and Localities URL: https://www.governing.com/gov-data/census/gentrification-in-cities-governing-report.html.

Manaugh, K., Miranda-Moreno, L.F., El-Geneidy, A.M., 2010. The effect of neighbourhood characteristics, accessibility, home–work location, and demographics on commuting distances. Transportation 37, 627–646.

Manville, M., Taylor, B.D., Blumenberg, E., 2018. Falling transit ridership: California and southern california .

McKinnish, T., Walsh, R., White, T.K., 2010. Who gentrifies low-income neighborhoods? Journal of urban economics 67, 180–193.

Mills, T., Steele, M., 2017. Transit center. URL: https://transitcenter.org/in-portland-economic-displacement-may-be-a-driver-of-transit-ridership-loss/.

Mucci, R.A., Erhardt, G.D., 2018. Evaluating the ability of transit direct ridership models to forecast medium-term ridership changes: Evidence from san francisco. Transportation Research Record , 0361198118758632.

Pasha, M., Rifaat, S.M., Tay, R., De Barros, A., 2016. Effects of street pattern, traffic, road infrastructure, socioeconomic and demographic characteristics on public transit ridership. KSCE Journal of Civil Engineering 20, 1017–1022.

Pathak, R., Wyczalkowski, C.K., Huang, X., 2017. Public transit access and the changing spatial distribution of poverty. Regional Science and Urban Economics 66, 198–212.

Phinney, R., 2013. Exploring residential mobility among low-income families. Social Service Review 87, 780–815.

Rayle, L., Dai, D., Chan, N., Cervero, R., Shaheen, S., 2016. Just a better taxi? a survey-based comparison of taxis, transit, and ridesourcing services in san francisco. Transport Policy 45, 168–178.

Richardson, J., Mitchell, B., Franco, J., 2019. Shifting neighborhoods: Gentrification and cultural displacement in American cities.

Rothstein, R., 2017. The color of law: A forgotten history of how our government segregated America. Liveright Publishing.

Schaller, B., 2018. The new automobility: Lyft, uber and the future of american cities .

Sikder, S., 2019. Who uses ride-hailing services in the united states? Transportation Research Record , 0361198119859302.

Slater, T., 2009. Missing marcuse: On gentrification and displacement. City 13, 292–311.

Squires, G.D., Kubrin, C.E., 2005. Privileged places: Race, uneven development and the geography of opportunity in urban america. Urban Studies 42, 47–68.

Taylor, B.D., Miller, D., Iseki, H., Fink, C., 2009. Nature and/or nurture? analyzing the determinants of transit ridership across us urbanized areas. Transportation Research Part A: Policy and Practice 43, 60–77.

Thompson, G.L., Brown, J.R., 2006. Explaining variation in transit ridership in us metropolitan areas between 1990 and 2000: multivariate analysis. Transportation Research Record 1986, 172–181.

Walls, M., Safirova, E., Jiang, Y., 2007. What drives telecommuting? relative impact of worker demographics, employer characteristics, and job types. Transportation Research Record 2010, 111–120.

Watkins, K., Berrebi, S., Diffee, C., Kiriazes, B., Ederer, D., 2019. Analysis of recent public transit ridership trends. TCRP Research Report 209 doi:`10.17226/25635`.

Wooldridge, J.M., 2002. Econometric analysis of cross section and panel data.

Zook, M., Shelton, T., Poorthuis, A., 2017. Big data and the city .